\documentclass{ifacconf}

\usepackage{graphicx}      
\usepackage{natbib}        

\usepackage[utf8]{inputenc} 
\usepackage[T1]{fontenc}    
\usepackage{url}            
\usepackage{booktabs}       
\usepackage{amsfonts}       
\usepackage{nicefrac}       
\usepackage{microtype}      

\usepackage{amsmath,amsfonts,bm}

\usepackage{comment}
\usepackage{subcaption}
\usepackage{caption}

\usepackage{bbm}

\usepackage[ruled,vlined,algo2e]{algorithm2e}

\newtheorem{Theorem}{Theorem}

\newtheorem{remark}{Remark}

\begin{document}
\begin{frontmatter}

\title{Opportunities and Challenges from Using Animal Videos in Reinforcement Learning for Navigation~\thanksref{footnoteinfo}} 
\thanks[footnoteinfo]{Research partially supported by the NSF under grants CCF-2200052, IIS-1914792, and DMS-1664644, by the ONR under grants N00014-19-1-2571 and N00014-21-1-2844, by the NIH under grants R01 GM135930 and UL54 TR004130, and by the Boston University Kilachand Fund for Integrated Life Science and Engineering.}

\author[First]{Vittorio Giammarino} 
\author[First]{James Queeney} 
\author[Second]{Lucas C. Carstensen} 
\author[Second]{Michael E. Hasselmo} 
\author[Third]{Ioannis Ch. Paschalidis} 

\address[First]{Division of Systems Engineering, Boston University, Boston, MA 02215, USA,  E-mail: {\tt \{vgiammar, jqueeney\}@bu.edu}.}
\address[Second]{Center for Systems Neuroscience, Department of Psychological and Brain Sciences and Graduate Program for Neuroscience, Boston University, Boston, MA 02215 USA, E-mail: {\tt \{lucasc, hasselmo\}@bu.edu}.}
\address[Third]{College of Engineering and Hariri Institute for Computing \& Computational Science \& Engineering, Boston University, Boston, MA 02215, USA, E-mail: {\tt yannisp@bu.edu}.}

\begin{abstract}                
We investigate the use of animal videos (observations) to improve Reinforcement Learning (RL) efficiency and performance in navigation tasks with sparse rewards. 
Motivated by theoretical considerations, we make use of weighted policy optimization for off-policy RL and describe the main challenges when learning from animal videos. We propose solutions and test our ideas on a series of 2D navigation tasks. We show how our methods can leverage animal videos to improve performance over RL algorithms that do not leverage such observations.
\end{abstract}


\end{frontmatter}

\section{Introduction}
Reinforcement Learning (RL) is a powerful tool to learn control policies from experience where the learning agent does not have explicit information about the task but is only allowed to interact with the environment and observe rewards \citep{sutton2018reinforcement}. Unfortunately, the trial-and-error nature of RL is often an obstacle for its actual deployment in real-world scenarios \citep{dulac2019challenges}. A solution to this problem is offered by the offline RL framework \citep{fu2020d4rl}, where control policies are learned entirely from static datasets without further interaction with the environment.

In realistic scenarios, leveraging data collected offline becomes a much more difficult task for the following reasons. First, datasets are not always collected in a demonstration format, that is, sequences of states, actions, and reward tuples. Instead, datasets are often collected in an observation format, i.e., sequences of states only, preventing the use of pure offline algorithms. Even when states and actions are collectable, the presence of domain differences between the source, i.e., the agent from which we collect data, and the target agent (the learner) limits the usefulness of the data. Finally, no information about the data collection policy is available and therefore techniques for off-policy corrections, such as Importance Sampling (IS) (refer to \citealt{queeney2021generalized}), become unusable.  

In this paper, we leverage videos of animals navigating an environment as offline datasets. We propose a hybrid approach that uses learner interactions to extract ready-to-use demonstrations from videos and uses the demonstrations to improve RL efficiency and performance. In order to accomplish our goal, we address three main problems throughout the paper: $(i)$ dealing with Domain Adaptation (DA) issues, $(ii)$ dealing with videos as offline datasets, and $(iii)$ theoretically motivating off-policy learning algorithms.

\textbf{Contributions:} We start by formulating an adversarial DA algorithm \citep{tzeng2017adversarial} that maps both the offline dataset and the online interactions of the learning agent into a common space, allowing for the use of data from different domains. Further, we outline a simple method based on supervised learning to recover demonstrations (states, actions, and rewards) from observations (states only). Finally, we propose an IS-free off-policy algorithm called {\em Advantage Weighted Policy Optimization (AWPO)} and discuss its connection to weighted regression RL algorithms such as Relative Entropy Policy Search (REPS) \citep{peters2010relative}, Maximum a posteriori Policy Optimization (MPO) \citep{abdolmaleki2018maximum}, Advantage Weighted Regression (AWR) \citep{peng2019advantage}, and Advantage Weighted Actor Critic (AWAC) \citep{nair2020awac} in the light of a recent work on principled off-policy learning \citep{queeney2021generalized}.

\textbf{Notation:} Unless indicated otherwise, we use uppercase letters (e.g., $S_t$) for random variables, lowercase letters (e.g., $s_t$) for values of random variables, script letters (e.g., $\mathcal{S}$) for sets, and bold lowercase letters (e.g., $\bm{\theta}$) for vectors. Let $[t_1 : t_2]$ be the set of integers $t$ such that $t_1 \leq t \leq t_2$; we write $S_t$ such that $t_1 \leq t \leq t_2$ as $S_{t_1 : t_2}$. We denote with $\mathbb{E}[\cdot]$ expectation, with $\mathbb{P}(\cdot)$ probability, and with $\mathbbm{1}_{S}(s)$ the indicator function which is equal to 1 when $S=s$ and 0 otherwise. $\mathcal{N}(\mu, \sigma)$ denotes a normal distribution with mean equal to $\mu$  and standard deviation equal to $\sigma$. $\mathbb{D}_{\text{KL}}(\cdot||\cdot)$ denotes the Kullback-Leibler (KL) divergence and $\mathbb{D}_{\text{TV}}(\cdot,\cdot)$ the total variation (TV) distance.


\section{Preliminaries}
We consider an infinite-horizon discounted Markov Decision Process (MDP) defined by the tuple $(\mathcal{S}, \mathcal{A}, P, r, d_0, \gamma)$ where $\mathcal{S}$ is the set of states and $\mathcal{A}$ is the set of actions, both possibly infinite. $P:\mathcal{S}\times \mathcal{A} \rightarrow \Delta_{\mathcal{S}}$ is the transition probability function and $\Delta_{\mathcal{S}}$ denotes the space of probability distributions over $\mathcal{S}$. The function $r:\mathcal{S}\times \mathcal{A} \rightarrow \mathbb{R}$ maps state-action pairs to rewards. $d_0\in\Delta_{\mathcal{S}}$ is the initial state distribution and $\gamma \in [0,1)$ the discount factor. We model the decision agent as a stationary policy $\pi:\mathcal{S}\rightarrow\Delta_{\mathcal{A}}$, where $\pi(a|s)$ is the probability of taking action $a$ in state $s$. The goal is to choose a policy that maximizes the expected total discounted reward $J(\pi)=\mathbb{E}_{\tau \sim \pi}[\sum_{t=0}^{\infty}\gamma^t r(s_t,a_t)]$, where $\tau = (s_0,a_0,s_1,a_1,\dots)$ is a trajectory sampled according to $s_0 \sim d_0$, $a_t\sim\pi(\cdot|s_t)$ and $s_{t+1}\sim P(\cdot|s_t,a_t)$. A policy $\pi$ induces a normalized discounted state visitation distribution $d^{\pi}$, where $d^{\pi}(s) = (1-\gamma)\sum_{t=0}^{\infty}\gamma^t\mathbb{P}(s_t=s | d_0,\pi,P)$. We define the corresponding normalized discounted state-action visitation distribution as $\mu^{\pi}(s,a)= d^{\pi}(s)\pi(a|s)$. We denote the state value function of $\pi$ as $V^{\pi}(s) = \mathbb{E}_{\tau \sim \pi}[\sum_{t=0}^{\infty}\gamma^t r(s_t,a_t)|S_0=s]$, the state-action value function as $Q^{\pi}(s,a) = \mathbb{E}_{\tau \sim \pi}[\sum_{t=0}^{\infty}\gamma^t r(s_t,a_t)|S_0=s, A_0=a]$ and the advantage function as $A^{\pi}(s,a) = Q^{\pi}(s,a) - V^{\pi}(s)$. Finally, when a policy is parameterized with parameters $\bm{\theta} \in \varTheta \subset \mathbb{R}^k$ we write $\pi_{\bm{\theta}}$.

\section{Domain adaptation and estimation of actions-rewards from videos}

In this section we formalize the concepts of domain adaptation and actions-rewards estimation \citep{higgins2017darla, schmeckpeper2020reinforcement, xing2021domain}; we present ways to address them within the RL framework.

\begin{figure*}[h!]
    \centering
    \begin{subfigure}[t]{0.3\textwidth}
        \centering
        \includegraphics[width=4cm, height=4cm]{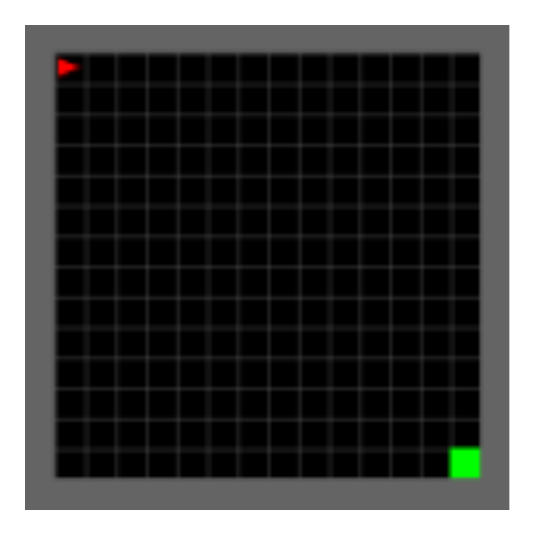}
        \caption{}
        \label{fig:human}
    \end{subfigure}
    ~
    \begin{subfigure}[t]{0.3\textwidth}
        \centering
        \includegraphics[width=4cm, height=4cm]{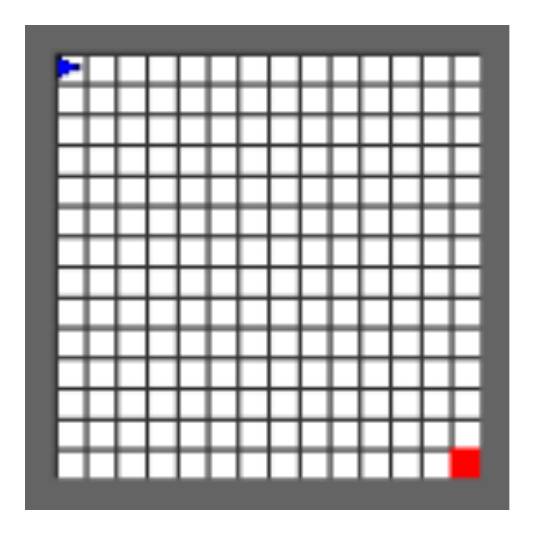}
        \caption{}
        \label{fig:human_modified}
    \end{subfigure}
    ~
    \begin{subfigure}[t]{0.3\textwidth}
        \centering
        \includegraphics[width=4cm, height=4cm]{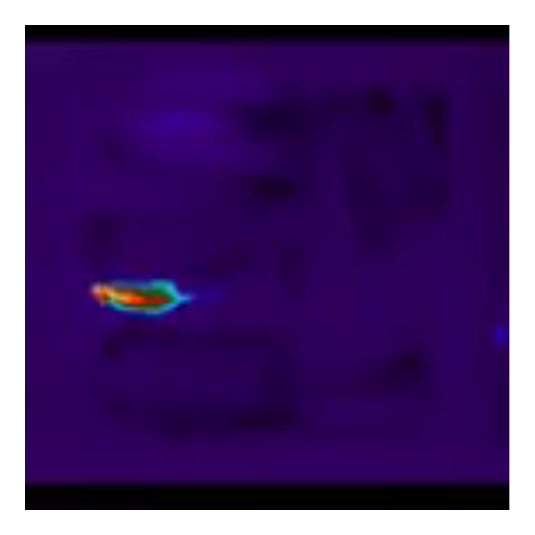}
        \caption{}
        \label{fig:rodent}
    \end{subfigure}
    \caption{MDPs used in the RL experiments. For all the MDPs the state space is made of $128 \times 128$ RGB images showing the top view of the environment. The action spaces are the navigation primitives north, south, east and west and the reward is $+1$ when a point in the bottom right corner is reached and a $0.001$ penalty for each decision step. Fig.~\ref{fig:human} shows the target MDP. Fig.~\ref{fig:human_modified} and Fig.~\ref{fig:rodent} show two different source MDPs from which we collect videos used as offline datasets in Algorithm~\ref{alg:AWPO}. A more extensive description of the rodent video in Fig.~\ref{fig:rodent} is provided in Fig.~\ref{fig:rodent_video}. For both the environments in Fig.~\ref{fig:human} and Fig.~\ref{fig:human_modified}, we rely on the open-source library minigrid \citep{minigrid}.}
    \label{fig:isomorphism}
\end{figure*}

We denote the source and the target domains as $D_S$ and $D_T$, respectively. Each domain corresponds to an MDP: $D_S \equiv (\mathcal{S}_S, \mathcal{A}_S, P_S, r_S, d_{0}^S, \gamma)$ and $D_T \equiv (\mathcal{S}_T, \mathcal{A}_T, P_T, r_T, d_{0}^T, \gamma)$ where $\gamma$ is a common discount factor. $\mathcal{S}$ and $d_0$ are different between the source and target domains, while $\mathcal{A}$ is shared and $P$ and $r$ have structural similarities. Refer to Fig.~\ref{fig:isomorphism} for an example. Fig.~\ref{fig:human} shows our target $D_T$ and Fig.~\ref{fig:human_modified} and \ref{fig:rodent} show two possible sources $D_S$. The state spaces, made of RGB pixels, show significant differences between $D_S$ and $D_T$ ($\mathcal{S}_S \neq \mathcal{S}_T$). On the other hand, all the domains share the same action spaces ($\mathcal{A}_S = \mathcal{A}_T$), since the control action can be modeled using the same motion primitives: north, south, east, west. Finally, transition probabilities and rewards have similarities ($P_S\approx P_T$ and $r_S \approx r_T$) since all the domains are governed by the same deterministic transition dynamics and the performance depends on the position of the agent with respect to a point in the environment.
In every source domain $D_S$, we can record a video of a quasi-optimal agent while performing the task. We refer to this video as the offline dataset $(s_S, s_S')_{1:M}$ made only of observations. Our goal is leveraging $(s_S, s_S')_{1:M}$ to achieve faster policy learning in $D_T$. 

Typically, in deep RL from pixels, agents learn an end-to-end mapping from states $s_T \in \mathcal{S}_T$ to actions $a_T \in \mathcal{A}_T$ \citep{mnih2013playing, mnih2015human}. In the process, a function $f_T: \mathcal{S}_T \to \mathcal{H}$ that maps the high-dimensional pixel space into a lower dimensional embedding space $\mathcal{H}$ is implicitly learnt and a policy $\pi_{\bm{\theta}}:\mathcal{H} \to \Delta_{\mathcal{A}}$ learns the mapping from $\mathcal{H}$ to $\Delta_{\mathcal{A}}$. In order to leverage $(s_S, s_S')_{1:M}$ in $D_T$, we define, in addition to $f_T$, another encoding function $f_S: \mathcal{S}_S \to \mathcal{H}$. Our domain adaptation step will focus on training $f_S$ such that the features extracted from $\mathcal{S}_S$ by $f_S$ will match as much as possible those extracted from $\mathcal{S}_T$ by $f_T$. When this is accomplished, the policy $\pi_{\bm{\theta}}$ can be directly trained on the embedding space $\mathcal{H}$ leveraging data collected from both $D_T$ and $D_S$.

Furthermore, given $(s_S, s_S')_{1:M}$ and the two encoders $f_T$ and $f_S$, our actions-rewards estimation will consist in training two inverse models $f^{\text{im}}_{\bm{\chi}_a}: \mathcal{H} \times \mathcal{H} \rightarrow \Delta_{\mathcal{A}}$ and $f^{\text{im}}_{\bm{\chi}_r}: \mathcal{H} \times \mathcal{H} \rightarrow \mathbb{R}$ using $(h_T, a_T, r_T, h_T') \equiv (f_T(s_T), a_T, r_T, f_T(s_T'))$ in order to estimate actions and rewards for $(h_S, h_S')_{1:M} \equiv (f_S(s_S), f_S(s_S'))_{1:M}$.
The full set $(h_T, a_T, r_T, h_T') \cup (h_S, \hat{a}_S, \hat{r}_S, h_S')$ will be then used to train $\pi_{\bm{\theta}}$.

\subsection{Domain adaptation}
We define two different encoding functions $f^{\text{enc}}_{\bm{\delta}_T}: \mathcal{S}_T \rightarrow \mathcal{H}$ and $f^{\text{enc}}_{\bm{\delta}_S}: \mathcal{S}_S \rightarrow \mathcal{H}$. Since we are interested in learning from videos, both $f^{\text{enc}}_{\bm{\delta}_T}$ and $f^{\text{enc}}_{\bm{\delta}_S}$ are parameterized with Convolutional Neural Networks (CNN) which take RGB images as input and respectively output $h_T \in \mathcal{H}$ and $h_S \in \mathcal{H}$. We train both $f^{\text{enc}}_{\bm{\delta}_T}$ and $f^{\text{enc}}_{\bm{\delta}_S}$ end-to-end with a value function $V_{\bm{\theta}_V}:\mathcal{H} \to \mathbb{R}$ or a state-action value function $Q_{\bm{\theta}_Q}:\mathcal{H} \times \mathcal{A} \to \mathbb{R}$ (see Algorithm~\ref{alg:AWPO}). Additionally, we train $f^{\text{enc}}_{\bm{\delta}_S}$ using adversarial DA as in \cite{tzeng2017adversarial}. Our adversarial DA step is formulated as the following $\min$-$\max$ game
\begin{align}
    \min_{\bm{\delta}_S} \max_{\bm{\phi}} \frac{1}{m}\sum_{i=1}^m[f^{\text{disc}}_{\bm{\phi}}(f^{\text{enc}}_{\bm{\delta}_T}(s_T^{(i)})) -  f^{\text{disc}}_{\bm{\phi}}(f^{\text{enc}}_{\bm{\delta}_S}(s_S^{(i)}))],
    \label{eq:adv_loss}
\end{align}
where $\{s_S^{(i)}\}_{i=1}^m \sim (s_S, s_S')_{1:M}$, $\{s_T^{(i)}\}_{i=1}^m \sim d^{\pi_{\bm{\theta}_k}}$ and $f^{\text{disc}}_{\bm{\phi}}:\mathcal{H}\to \mathbb{R}$ is a discriminator function which recognizes from which domain ($D_S$ or $D_T$) the embedding has been generated. $f^{\text{enc}}_{\bm{\delta}_S}$ is trained in order to make the features $h_S$ as close as possible to $h_T$. At optimality, the discriminator is no longer able to distinguish from which MDP the embedding $h \in \mathcal{H}$ has been generated, meaning that both $\mathcal{S}_S$ and $\mathcal{S}_T$ have been mapped into the same common space $\mathcal{H}$. Note that in \eqref{eq:adv_loss} we use the Wasserstein loss as introduced in \cite{arjovsky2017wasserstein}. We prefer this loss to the original binary cross-entropy in \cite{goodfellow2014generative} for stability reasons.

\subsection{Actions and rewards estimation}
Recall $h_S \equiv f^{\text{enc}}_{\bm{\delta}_S}(s_S)$, $h_T \equiv f^{\text{enc}}_{\bm{\delta}_T}(s_T)$ and the source data $(s_S, s_S')_{1:M}$ which do not contain actions $(a_S)_{1:M}$ and rewards $(r_S)_{1:M}$. A popular approach in the literature is to estimate $(a_S)_{1:M}$ and $(r_S)_{1:M}$ using inverse models, where the online interactions $(h_T, a_T, r_T, h_T')$ collected by the learner in $D_T$ are used for supervised learning \citep{torabi2018behavioral, schmeckpeper2020reinforcement, pathak2017curiosity}. We define two inverse models $f^{\text{im}}_{\bm{\chi}_a}: \mathcal{H} \times \mathcal{H} \rightarrow \Delta_{\mathcal{A}}$ and $f^{\text{im}}_{\bm{\chi}_r}: \mathcal{H} \times \mathcal{H} \rightarrow \mathbb{R}$ for actions and rewards, respectively. We collect $(h_T, a_T, r_T, h_T')$ in $D_T$ and train the two inverse models to minimize the following losses
\begin{align}
\begin{split}
    \mathcal{L}^{\text{im}}_a(\bm{\chi}_a) &= ||p(a_T) - f^{\text{im}}_{\bm{\chi}_a}(h_T,h_T')||^2, \\
    \mathcal{L}^{\text{im}}_r(\bm{\chi}_r) &= (r_T - f^{\text{im}}_{\bm{\chi}_r}(h_T,h_T'))^2,
\end{split}
\label{eq:actions-rewards-loss}
\end{align}
where $p(a_T)$ is a distribution over $\mathcal{A}$. In practice, when $\mathcal{A}$ is a discrete action space $p(a_T) = \mathbbm{1}_{A_T}(a_T)$, and when $\mathcal{A}$ is continuous $p(a_T) = \mathcal{N}(a_T, \sigma)$ with $\sigma$ arbitrary.
We then sample $\hat{a}_S \sim f^{\text{im}}_{\bm{\chi}_a}(h_S,h_S')$ and compute $\hat{r}_S = f^{\text{im}}_{\bm{\chi}_r}(h_S,h_S')$ obtaining the tuples $(h_S, \hat{a}_S, \hat{r}_S, h_S')_{1:M}$ which are combined with $(h_T, a_T, r_T, h_T')$ for policy learning in $D_T$.
Finally, because we assume that the demonstrating agent in $D_S$ is quasi-optimal, we add a small constant bonus $r_i$ to $\hat{r}_S$ to encourage the learner to follow the demonstrator and compensate for reward sparsity.

\section{Advantage Weighted Policy Optimization}
The final challenge in order to successfully leverage videos of animals in RL is determining the proper algorithm able to learn off-policy. Recall $\pi_{\bm{\theta}}$ is defined as $\pi_{\bm{\theta}}:\mathcal{H} \to \Delta_{\mathcal{A}}$; we therefore consider $\mathcal{H}$ as our state space of interest. In the following sections we state instrumental results for our derivations.

\textbf{Policy improvement lower bound:} Our starting point is the policy improvement lower bound originally developed in \cite{kakade2002approximately} and refined in \cite{achiam2017constrained}.
\begin{Theorem}[\citealt{achiam2017constrained}]
Consider the current policy $\pi_k$. For any future policy $\pi$ we have
\begin{align}
    \begin{split}
            J(\pi) - J(\pi_k) \geq& \frac{1}{1-\gamma} \mathbb{E}_{(h,a) \sim \mu^{\pi_k}}\bigg[\frac{\pi(a|h)}{\pi_k(a|h)}A^{\pi_k}(h,a)\bigg] \\
            &- \frac{2\gamma C^{\pi,\pi_k}}{(1-\gamma)^2}\mathbb{E}_{h\sim d^{\pi_k}}[\mathbb{D}_{\text{\normalfont TV}}(\pi_k(\cdot|h), \pi(\cdot|h))],
    \label{eq:achiam}
    \end{split}
\end{align}
where $C^{\pi,\pi_k} = \max_{h\in \mathcal{H}} |\mathbb{E}_{a\sim \pi(\cdot|h)}[A^{\pi_k}(h,a)]|$ and \\ $\mathbb{D}_{\text{\normalfont TV}}(\pi_k(\cdot|h), \pi(\cdot|h)) = \frac{1}{2}\int_{a \in \mathcal{A}}|\pi_k(a|h) - \pi(a|h)| da$ is the total variation distance between the distributions $\pi_k(\cdot|h)$ and $\pi(\cdot|h)$.
\label{Theo:Achiam}
\end{Theorem}

Theorem~\ref{Theo:Achiam} has been further generalized for the off-policy setting in \cite{queeney2021generalized} in order to allow for expectations with respect to any behavioral policy. 
\begin{Theorem}[\citealt{queeney2021generalized}]
Consider the current policy $\pi_k$ and the behavioral policy $\beta$. For any future policy $\pi$ we have
\begin{equation}
    \begin{split}
        J(\pi) - J(\pi_k) \geq& \frac{1}{1-\gamma}\mathbb{E}_{(h,a) \sim \mu^{\beta}}\bigg[\frac{\pi(a|h)}{\beta(a|h)}A^{\pi_k}(h,a)\bigg] \\
        &- \frac{2\gamma C^{\pi, \pi_k}}{(1-\gamma)^2}\mathbb{E}_{h \sim d^{\beta}}[\mathbb{D}_{\text{\normalfont TV}}(\beta(\cdot|h), \pi(\cdot|h))],
        \label{eq:lemma_GePPO}
    \end{split}
\end{equation}
where $C^{\pi,\pi_k}$ and $\mathbb{D}_{\text{\normalfont TV}}(\beta(\cdot|h), \pi(\cdot|h))$ are defined as in Theorem~\ref{Theo:Achiam}.
\label{Theo:Queeney}
\end{Theorem}

The first term on the right-hand side in both \eqref{eq:achiam} and \eqref{eq:lemma_GePPO} is referred to as the surrogate objective, while the second is a penalty term or a soft constraint. Note that Theorem~\ref{Theo:Achiam} and Theorem~\ref{Theo:Queeney} show that by enforcing any future policy $\pi$ to stay close to $\pi_k$ or $\beta$, a monotonic policy improvement can be guaranteed. For practical purposes, the following surrogate optimization problem for Theorem~\ref{Theo:Achiam} is derived in \cite{schulman2015trust}:
\begin{equation}
\begin{split}
    \max_{\pi} \ \ \ & \mathbb{E}_{(h,a) \sim \mu^{\pi_k}}\bigg[\frac{\pi(a|h)}{\pi_k(a|h)}A^{\pi_k}(h,a)\bigg] \\
    \text{s.t.} \ \ \ &\mathbb{E}_{h \sim d^{\pi_k}}[\mathbb{D}_{\text{KL}}(\pi_k(\cdot|h) || \pi(\cdot|h))] \leq \epsilon.
    \label{eq:TRPO_surrogate_problem}
\end{split}
\end{equation}
Based on \eqref{eq:TRPO_surrogate_problem}, the popular on-policy Proximal Policy Optimization (PPO) \citep{schulman2017proximal} and its off-policy version Generalized PPO (GePPO) \citep{queeney2021generalized} are formalized. \\

\textbf{AWPO derivations:} In the following, we focus on deriving from Theorem~\ref{Theo:Queeney} an off-policy algorithm that we call Advantage Weighted Policy Optimization (AWPO). This belongs to the family of weighted regression algorithms and has strong similarities with REPS, MPO, AWR and AWAC. The main source of difference lies in the way the algorithm is derived and in the need for an off-policy advantage estimation.
We start from the lower bound in Theorem~\ref{Theo:Queeney} and proceed analogously to \cite{achiam2017constrained}. Note that the total variation distance is a metric \citep{tsybakov1997nonparametric}, hence $\mathbb{D}_{\text{TV}}(\beta(\cdot|h), \pi(\cdot|h)) = \mathbb{D}_{\text{TV}}(\pi(\cdot|h), \beta(\cdot|h))$. By applying Pinsker's inequality followed by Jensen's inequality, we see that
\begin{equation*}
    \begin{split}
        &\mathbb{E}_{h \sim d^{\beta}}[\mathbb{D}_{\text{TV}}(\pi(\cdot|h), \beta(\cdot|h))] \\
        & \qquad \qquad \qquad \leq \mathbb{E}_{h \sim d^{\beta}}\bigg[\sqrt{\frac{1}{2}\mathbb{D}_{\text{KL}}(\pi(\cdot|h)|| \beta(\cdot|h))}\bigg] \\
        & \qquad \qquad \qquad \leq \sqrt{\frac{1}{2}\mathbb{E}_{h \sim d^{\beta}}[\mathbb{D}_{\text{KL}}(\pi(\cdot|h)|| \beta(\cdot|h))]}.    \end{split}
\end{equation*}
The expression above leads to the surrogate problem
\begin{equation}
    \begin{split}
            \max_{\pi} \ \ \ & \mathbb{E}_{h \sim d^{\beta}}\bigg[\mathbb{E}_{a \sim \pi(\cdot|h)}\big[A^{\pi_k}(h,a)\big]\bigg] \\
            \text{s.t.} \ \ \ &\mathbb{E}_{h \sim d^{\beta}}[\mathbb{D}_{\text{KL}}(\pi(\cdot|h) || \beta(\cdot|h))] \leq \epsilon,
    \end{split}
    \label{eq:AWAC_surrogate}
\end{equation}
 where $\beta$ is any behavioral policy.
By converting the KL divergence in \eqref{eq:AWAC_surrogate} as a soft constraint and solving the Lagrangian problem in closed form, we obtain the following optimal solution
\begin{align*}
    \pi^*(a|h) &= \beta(a|h) \exp \bigg( \frac{A^{\pi_k}(h,a)}{\lambda} \bigg)\frac{1}{Z(h)},
\end{align*}
where $Z(h)$ is a normalizing factor ensuring $\pi^*(\cdot|h)$ is a probability distribution and $\lambda$ is a Lagrange multiplier.
Further, we project $\pi^*$ on the manifold of function approximations parameterized by $\bm{\theta}$ \citep{peng2019advantage}. Given $\pi_{\bm{\theta}_k}$ the policy parameterized by $\bm{\theta}_k$ at step $k$, for $k+1$ we obtain
\begin{equation}
    \bm{\theta}_{k+1} = \arg \max_{\bm{\theta}} \mathbb{E}_{(h,a) \sim \mu^{\beta}}\bigg[\exp \bigg( \frac{A^{\pi_{\bm{\theta}_{k}}}(h,a)}{\lambda} \bigg)\log \pi_{\bm{\theta}}(a|h)\bigg].
    \label{eq:AWPO}
\end{equation}
Note that the Lagrange multiplier $\lambda$ in \eqref{eq:AWPO} is provided as hyperparameter.  
\begin{remark}[Novelty]
    We show direct connection with the generalized policy improvement in Theorem~\ref{Theo:Queeney} which is a missing piece in all the aforementioned literature \citep{peters2010relative,abdolmaleki2018maximum, peng2019advantage, nair2020awac}. Moreover, a novel small nuance of our formulation is the need for off-policy policy evaluation since we make use of samples from $\beta$ to compute $A^{\pi_{\bm{\theta}_{k}}}$ for the current policy $\pi_{\bm{\theta}_{k}}$.
\end{remark}
\begin{remark}[Off-policy policy evaluation]
\label{remark_off_policy_eval}
    Off-policy policy evaluation is defined as leveraging data generated by a behavioral policy $\beta$ to evaluate the performance metric of the evaluation policy $\pi_{\bm{\theta}_{k}}$ (refer to \citealt{chandak2021universal}). Well-known algorithms for off-policy policy evaluation are Retrace in \cite{munos2016safe}, V-trace in \cite{espeholt2018impala} and UnO in \cite{chandak2021universal}. However, all of these require IS correction, which implies knowledge of $\beta$. When learning from videos of animals, $\beta$ is unknown and cannot be estimated using Imitation Learning as in \cite{kostrikov2021offline, giammarino2022learning} given the absence of actions. We consider two main algorithms for off-policy evaluation which do not perform IS correction: Peng Q-Lambda (PQL) \citep{peng1994incremental, kozuno2021revisiting} and Generalized Advantage Estimation (GAE) \citep{schulman2015high}. PQL makes use of an explicit correction mechanism for off-policy policy evaluation, while GAE is originally conceived as an on-policy algorithm which, despite not accounting for off-policy data, works well in practice. We refer to \cite{chandak2021universal,munos2016safe,kozuno2021revisiting} for additional details on policy evaluation with off-policy correction.  
\end{remark}
\textbf{Advantage estimation:} As mentioned in Remark~\ref{remark_off_policy_eval}, we perform advantage estimation using either GAE or PQL. GAE makes use of a parameterized value function $V_{\bm{\theta}_V}$, which is trained by collecting rollouts of length $N$ in $D_T$ and length $M$ in $D_S$ and minimizing $\mathcal{L}(\bm{\theta}_V, \bm{\delta}_T, \bm{\delta}_S)$ in \eqref{eq:loss_V}
\begin{align}
    \begin{split}
        \mathcal{L}_T(\bm{\theta}_V, \bm{\delta}_T) =& \frac{1}{N}\sum_{t=1}^N||\hat{V}_T^{(t)} - V_{\bm{\theta}_V}(f^{\text{enc}}_{\bm{\delta}_T}(s_T^{(t)}))||^2, \\
        \mathcal{L}_S(\bm{\theta}_V, \bm{\delta}_S) =& \frac{1}{M}\sum_{t=1}^M||\hat{V}_S^{(t)} - V_{\bm{\theta}_V}(f^{\text{enc}}_{\bm{\delta}_S}(s_S^{(t)}))||^2,\\
        \mathcal{L}(\bm{\theta}_V, \bm{\delta}_T, \bm{\delta}_S) =& \frac{1}{2}(\mathcal{L}_T(\bm{\theta}_V, \bm{\delta}_T) + \mathcal{L}_S(\bm{\theta}_V, \bm{\delta}_S)),
    \end{split}
    \label{eq:loss_V}
\end{align}
where $\hat{V}_T^{(t)} = \sum_{l=0}^N \gamma^l r_T^{(t+l)}$ and $\hat{V}_S^{(t)} = \sum_{l=0}^M \gamma^l \hat{r}_S^{(t+l)}$ are discounted sums of rewards. This approach for estimating the value function is called TD(1) or Monte Carlo \citep{sutton2018reinforcement}. Note that $f^{\text{enc}}_{\bm{\delta}_T}$ and $f^{\text{enc}}_{\bm{\delta}_S}$ are also trained to minimize $\mathcal{L}(\bm{\theta}_V, \bm{\delta}_T, \bm{\delta}_S)$ in \eqref{eq:loss_V}. On the other hand, PQL makes use of a parameterized state-action value function $Q_{\bm{\theta}_Q}$ and follows the update rule in \cite{peng1994incremental, kozuno2021revisiting}. As for GAE, we update $f^{\text{enc}}_{\bm{\delta}_T}$ and $f^{\text{enc}}_{\bm{\delta}_S}$ together with $Q_{\bm{\theta}_Q}$.

The full algorithm is summarized in Algorithm~\ref{alg:AWPO}, which combines domain adaptation, source actions-rewards estimation and AWPO.

\begin{algorithm2e}
\caption{AWPO from Videos with Adversarial Domain Adaptation}
\label{alg:AWPO}
Input: $(s_S, s_S')_{1:M}$, $\pi_{\bm{\theta}}$, $V_{\bm{\theta}_V}$ ($Q_{\bm{\theta}_Q}$),  $f^{\text{enc}}_{\bm{\delta}_S}$, $f^{\text{enc}}_{\bm{\delta}_T}$, $f^{\text{disc}}_{\bm{\phi}}$, $f^{\text{im}}_{\bm{\chi}_a}$, $f^{\text{im}}_{\bm{\chi}_r}$, $r_i$, $\alpha$, $\lambda$, $K$, $N$.  \\
\For{$k=0,\ldots,K-1$}{
Generate $(s_T, a_T, r_T, s_T')_{1:N}$ from interaction of $\pi_{\bm{\theta}_k}$ with the environment. \\
Perform adversarial DA as in \eqref{eq:adv_loss}, and\\
$\hat{a}_S$ and $\hat{r}_S$ estimation as in \eqref{eq:actions-rewards-loss}.\\
Set $\hat{r}_S =  f^{im}_{\bm{\chi}_r}(h_S, h_S') + r_i$. \\
Set $\mathcal{D}\equiv (h_S, \hat{a}_S, \hat{r}_S, h_S')_{1:M} \cup (h_T, a_T, r_T, h_T')_{1:N}$.\\
Perform $A^{\pi_{\bm{\theta}_{k}}}(h,a)$ estimation.\\
Update 
$V_{\bm{\theta}_V}$ ($Q_{\bm{\theta}_Q}$),  $f^{\text{enc}}_{\bm{\delta}_S}$, $f^{\text{enc}}_{\bm{\delta}_T}$ using \eqref{eq:loss_V} (or \citealt{peng1994incremental}). \\
$\bm{\theta}_{k+1} = \bm{\theta}_{k} + \alpha \nabla_{\bm{\theta}} \mathbb{E}_{\mathcal{D}}\bigg[\exp \bigg( \frac{A^{\pi_{\bm{\theta}_{k}}}(h,a)}{\lambda} \bigg)\log \pi_{\bm{\theta}}(a|h)\bigg].$\\
$\bm{\theta}_{k} \leftarrow \bm{\theta}_{k+1}.$
}
$\bm{\theta} \leftarrow \bm{\theta}_{K}.$\\
\Return $\pi_{\bm{\theta}}$
\end{algorithm2e}

\section{Experiments}
\begin{figure*}[h!]
    \centering
    \includegraphics[width=\linewidth]{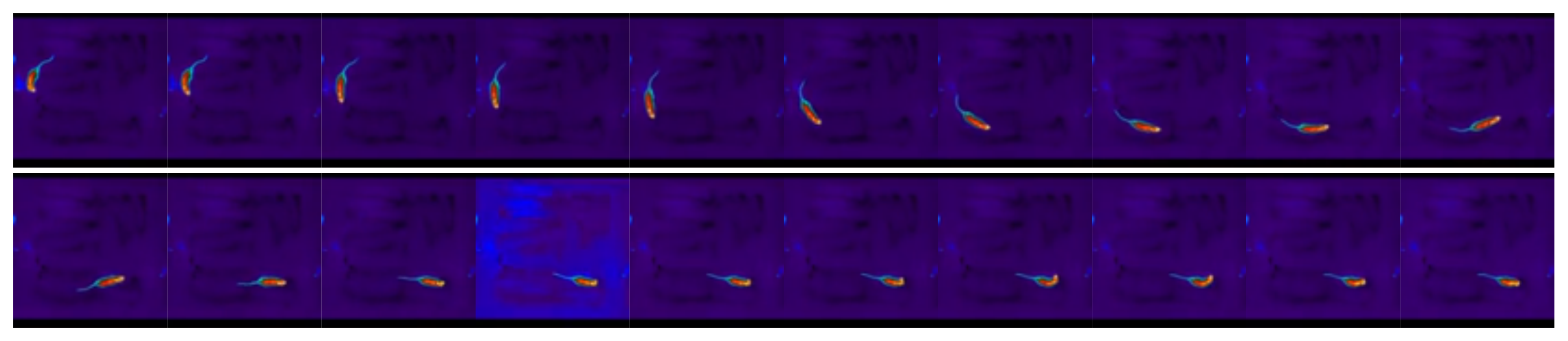}
    \caption{Frames from the video used as offline dataset. The video consists of a top view of a rodent foraging in an empty open field maze. The video is recorded using a thermal camera.}
    \label{fig:rodent_video}
\end{figure*}

We conduct a series of experiments in order to evaluate all components of our algorithm. We focus on 2D navigation tasks with sparse rewards from visual observations. We test Algorithm~\ref{alg:AWPO} in RL with offline observations without DA, with simplified DA (environment in Fig.~\ref{fig:human} as target and Fig.~\ref{fig:human_modified} as source for both these experiments), and finally with hard DA where the rodent video in Fig.~\ref{fig:rodent_video} is used as an offline dataset (Fig.~\ref{fig:human} for target and Fig.~\ref{fig:rodent} for source). Videos of rats foraging in an open-field arena were collected at 1024×1024 spatial resolution and 14-bit thermal resolution per pixel using thermal infrared cameras (FLIR SC8000, FLIR Systems, Inc.). The dataset used is a subset utilizing only the top-down view of the Rodent3D dataset, in which a full description is available at \cite{patel2022animal}. Finally, we parameterize the encoders using a CNN with 3 layers and the discriminator, the policy network, the value function, and the inverse models using multilayer perceptrons (MLP) with a single hidden layer. For additional implementation details we refer to the code publicly available at our GitHub repository\footnote{https://github.com/VittorioGiammarino/Opportunities-and-Challenges-of-Using-Animal-Videos-in-Reinforcement-Learning-for-Navigation.git}.

\subsection{Results}

\begin{figure}[h!]
    \centering
    \begin{subfigure}[t]{\linewidth}
        \centering
        \includegraphics[width=8.8cm, height=6cm]{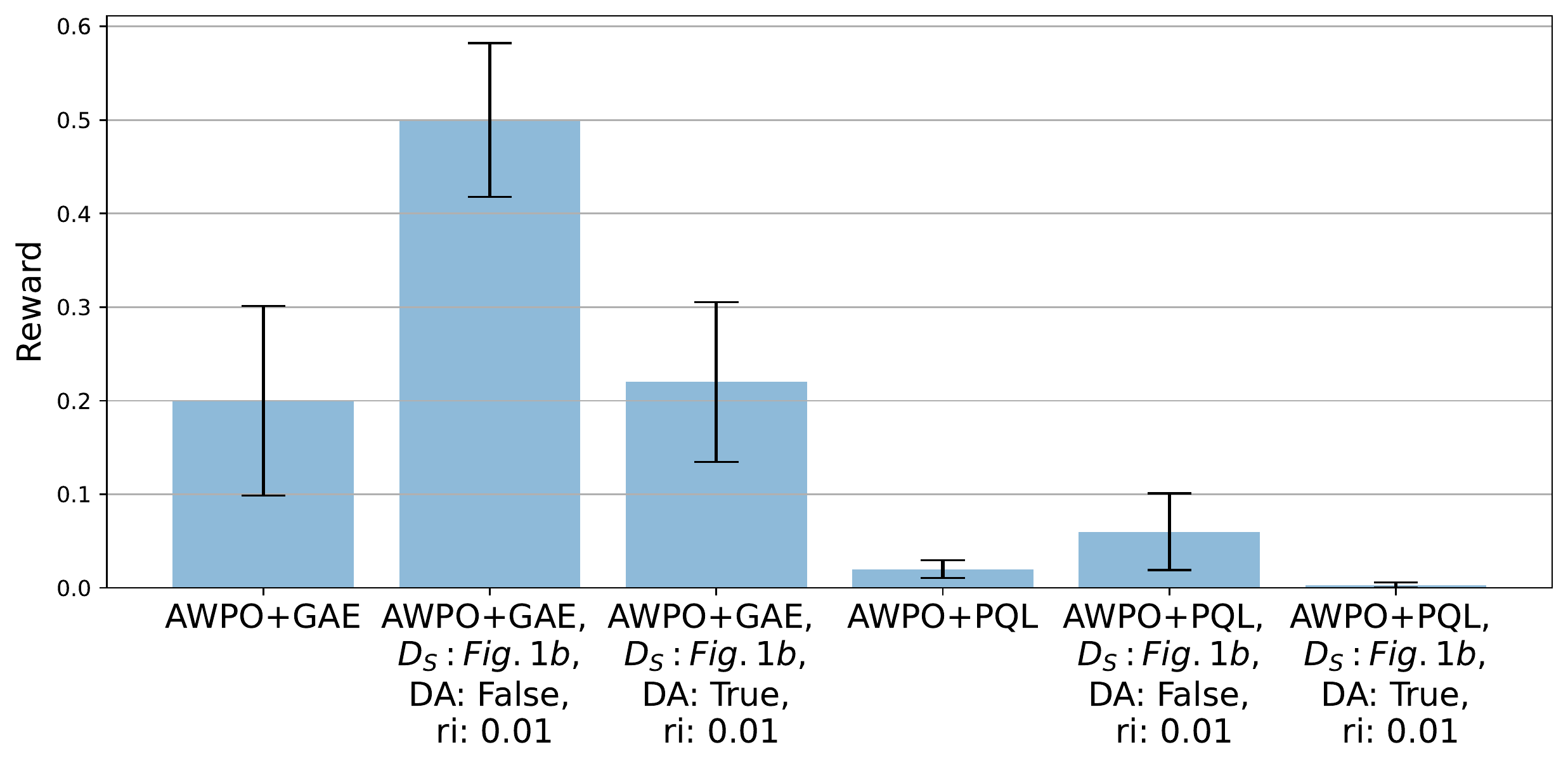}
        \caption{}
        \label{fig:results_human}
    \end{subfigure}
    ~
    \begin{subfigure}[t]{\linewidth}
        \centering
        \includegraphics[width=8.8cm, height=6cm]{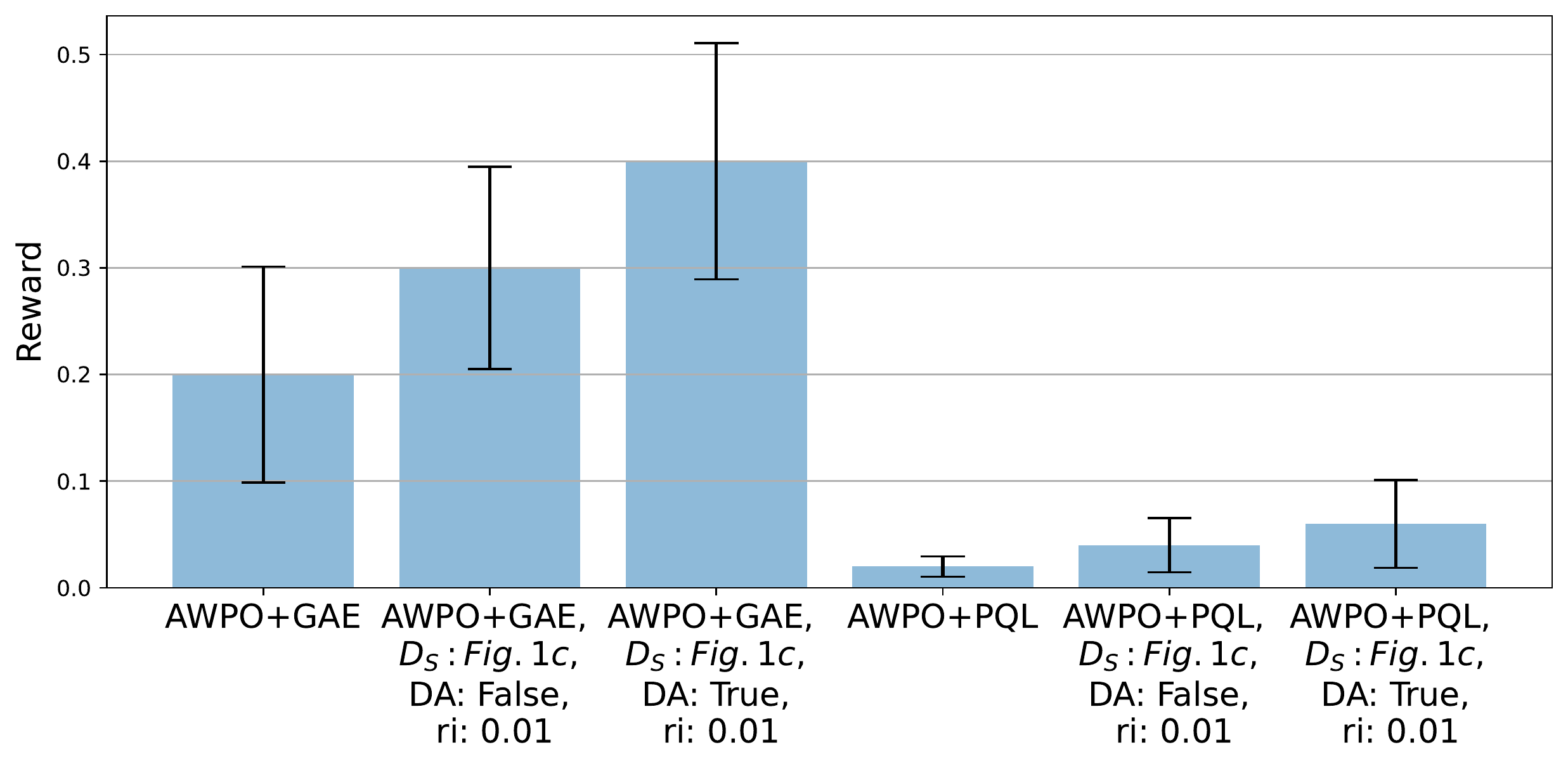}
        \caption{}
        \label{fig:results_rodent}
    \end{subfigure}
    \caption{Experiment results. The reported reward is averaged over the final 10 evaluations and 10 seeds. Error bars show the standard error. Each experiment consists of $400$k training steps. Both Fig.~\ref{fig:results_human} and Fig.~\ref{fig:results_rodent} contain the labels AWPO+GAE and AWPO+PQL which represent the RL-only benchmark where no use of offline datasets is made. The other labels indicate variations of Algorithm~\ref{alg:AWPO} where $D_S$ is the source domain in which the offline dataset is collected (either Fig.~\ref{fig:human_modified} or Fig.~\ref{fig:rodent}), DA indicates whether the DA step in \eqref{eq:adv_loss} is used or not and $r_i$ is the small bonus added to the estimated reward in the source domain.}
    \label{fig:results}
\end{figure}

We focus on the 2D navigation task in Fig.~\ref{fig:human}, where the state $s_T \in \mathcal{S}_T$ is the $128 \times 128$ RGB image showing the full grid and the goal is to reach the bottom right corner labeled by the green square. The agent receives $+1$ when reaching the goal and receives a $0.001$ penalty for each step. The episode starts with the agent in the top-left corner and the maximum episode length is set to $512$ steps. We run all the experiments for $400k$ training steps and evaluate the policy every $2k$ steps for $10$ episodes. This procedure is repeated for $10$ different random seeds. Fig.~\ref{fig:results} summarizes the final results where we report the reward obtained at the end of the training process averaged over episodes and seeds. In Fig.~\ref{fig:results}, AWPO+GAE and AWPO+PQL refer to pure RL algorithms where the agent learns only by its own experience. The other labels refer instead to Algorithm~\ref{alg:AWPO} where for $D_S=$\ref{fig:human_modified} we are leveraging a video collected in Fig.~\ref{fig:human_modified} as an offline dataset, while for $D_S=$\ref{fig:rodent} we are leveraging the rodent video in Fig.~\ref{fig:rodent_video}. Fig.~\ref{fig:results} shows that both versions of AWPO benefit from the introduction of videos as offline datasets and outperform their RL counterpart for both $D_S=$\ref{fig:human_modified} and $D_S=$\ref{fig:rodent}. We notice that our DA step is particularly useful when learning from the rodent (Fig.~\ref{fig:results_rodent}), while for $D_S=$\ref{fig:human_modified} this is not the case and our algorithm works better when the DA step is turned off. For $D_S=$\ref{fig:human_modified}, we notice $f^{\text{disc}}_{\bm{\phi}}$ is able to discriminate well between Fig.~\ref{fig:human} and Fig.~\ref{fig:human_modified} due to the background differences, making the DA problem in \eqref{eq:adv_loss} much more difficult for $f^{\text{enc}}_{\bm{\delta}_S}$. Improving the DA step in the context of policy learning will be the main focus of future works.

\section{Conclusions}
This paper analyzes the challenges of using videos in RL as offline datasets in order to increase learning efficiency and performance. The main problems we identify consist in dealing with domain discrepancies between source and target domains, learning from offline observations rather than demonstrations, and leveraging off-policy data for policy updates. We allow for structural differences in $\mathcal{S}$ between source and target domains, which we account for by applying domain adaptation techniques in order to make the offline dataset useful to the learning agent. The importance of this step is evident in Fig.~\ref{fig:results} for the $D_S=$\ref{fig:rodent} experiments. Furthermore, learning from offline observations deals with the problem of estimating actions and rewards from a static dataset consisting of observations only. Our solution is based on inverse models trained using supervised learning and learner's interactions. Finally, considering recent works on principled off-policy learning, we re-derive the AWPO algorithm and test it with and without offline videos on a 2D navigation task.
Future works will focus on improving each individual step with particular attention to DA which shows the greatest margin for improvement. Moreover, we plan to increase the scope of our experiments, focusing not only on the navigation setup, but also on other learning tasks such as locomotion and manipulation.

\bibliography{ifacconf}             

\end{document}